\documentclass[twocolumn,showpacs,preprintnumbers,nofootinbib,superscriptaddress]{revtex4}

\usepackage{graphicx}

\def\lsim{\raise0.3ex\hbox{$<$\kern-0.75em\raise-1.1ex\hbox{$\sim$}}}
\def\gsim{\raise0.3ex\hbox{$>$\kern-0.75em\raise-1.1ex\hbox{$\sim$}}}

\begin{document}

\preprint{UTHEP-569, TKYNT-08-17}


\title{Fixed Scale Approach to Equation of State in Lattice QCD} 


\author{T.~Umeda}\affiliation{Graduate School of Pure and Applied Sciences,
University of Tsukuba, Tsukuba, Ibaraki 305-8571, Japan}
\author{S.~Ejiri}\affiliation{Physics Department, Brookhaven National
Laboratory, Upton, New York 11973, USA}
\author{S.~Aoki}\affiliation{Graduate School of Pure and Applied Sciences,
University of Tsukuba, Tsukuba, Ibaraki 305-8571, Japan}
\affiliation{RIKEN BNL Research Center, Brookhaven National Laboratory, Upton, 
New York 11973, USA}
\author{T.~Hatsuda}\affiliation{Department of Physics, The University of Tokyo,
Tokyo 113-0033, Japan}
\author{K.~Kanaya}\affiliation{Graduate School of Pure and Applied Sciences,
University of Tsukuba, Tsukuba, Ibaraki 305-8571, Japan}
\author{Y.~Maezawa}\affiliation{En'yo Radiation Laboratory, Nishina
Accelerator Research Center, RIKEN, Wako, Saitama 351-0198, Japan} 
\author{H.~Ohno}\affiliation{Graduate School of Pure and Applied Sciences,
University of Tsukuba, Tsukuba, Ibaraki 305-8571, Japan}
\collaboration{WHOT-QCD Collaboration}

\date{\today}


\begin{abstract}
A new approach to study the equation of state in finite-temperature QCD 
is proposed on the lattice.
Unlike the conventional method in which
the temporal lattice size $N_t$ is fixed,
the temperature $T$  
is varied by changing $N_t$ at fixed lattice scale.
The pressure of the hot QCD plasma is calculated by 
the integration of the trace anomaly with respect to $T$
at fixed lattice scale.
This ``$T$-integral method''
is tested in quenched QCD on isotropic and anisotropic lattices
and is shown to give reliable results especially at intermediate and
low temperatures. 
\end{abstract}

\maketitle

\section{Introduction}

The equation of state (EOS) is one of the most fundamental observables 
to identify different phases in 
 quantum chromodynamics (QCD) at finite temperature $T$. Also,
it is an essential input to describe the
space-time evolution of the hot QCD matter
created in relativistic heavy ion collisions \cite{Hirano:2008hy}.
So far, the lattice QCD is the only systematic method to calculate EOS
for wide range of $T$ across the region of the hadron-quark phase
transition (see the recent review, \cite{lat07}).  

Conventionally, the EOS on the lattice is extracted from a method in
which $T=(N_t a)^{-1}$ is varied by changing the lattice scale $a$ (or
equivalently the lattice gauge coupling $\beta = 6/g^{2}$) with a
fixed temporal lattice size $N_t$. 
Using the thermodynamic relation $p = (T/V) \ln Z$ 
with $V$ being the spatial volume and $Z$ being the partition
function, the pressure $p$ is calculated as \cite{inte_method} 
\begin{equation}
p = \frac{T}{V} \int^{\beta}_{\beta_0} \! d\beta \, \frac{1}{Z}
\frac{\partial Z}{\partial \beta} 
= -\frac{T}{V} \int^{\beta}_{\beta_0} \! d\beta \left\langle
\frac{\partial S}{\partial \beta} \right\rangle . 
\end{equation}
Here $S$ is the lattice action and $\langle \cdots \rangle$ is the
thermal average  
with zero temperature contribution subtracted. 
(In multi-parameter cases such as QCD with dynamical quarks,
``$\beta$'' should be  
generalized to the position vector in the coupling parameter space \cite 
{AliKhan:2001ek}.) 
The initial point of integration $\beta_0$ is chosen in the low
temperature phase from the condition 
$p(\beta_0) \approx 0$.

In this conventional method, major part of the computational cost
is devoted to zero temperature simulations; they
are necessary to set the lattice scale and 
to carry out zero-temperature subtractions at the simulation points.
Furthermore, for the calculation of the trace anomaly 
$\epsilon-3p$ with $\epsilon$ being the energy density, 
the non-perturbative beta functions have to be determined by zero
temperature simulations at the same simulation points. 
In multi-parameter cases, simulations should be performed on a line of
constant physics (LCP) in the coupling parameter space in order to
identify the effect of temperature  
to a given physical system. LCP's should be determined also at $T=0$.
These zero temperature simulations for a wide range of coupling
parameters are 
numerically demanding, in particular, for QCD with dynamical quarks.

In this paper, we push an alternative approach where
temperature is varied by $N_t$ with other parameters fixed. 
The fixed scale approach has been applied with the derivative method
\cite{Engels:1980ty} in, e.g. Ref.~\cite{Levkova:2006gn}.
However, determination of nonperturbative Karsch coefficients for all
the simulation points requires quite a work, which will not be easy
for QCD with dynamical quarks. 
Here, we propose a new method, ``the $T$-integral method'':
To calculate the pressure non-perturbatively, 
we use 
\begin{eqnarray}
\frac{p}{T^4} = \int^{T}_{T_0} dT \, \frac{\epsilon - 3p}{T^5},
\label{eq:1}
\end{eqnarray}
which is obtained from the thermodynamic relation valid
at vanishing chemical potential:
\begin{eqnarray}
T \frac{\partial}{\partial T} \left( \frac{p}{T^4} \right) =
\frac{\epsilon-3p}{T^4}. 
\end{eqnarray}
The initial temperature $T_0$ is chosen such that $p(T_0) \approx 0$.
Calculation of $\epsilon -3p$ requires the beta functions just at the
simulation point, but no further Karsch coefficients are necessary. 
Since $T$ is restricted to have discrete values,
we need to make an interpolation of $(\epsilon -3p)/T^4$ with respect
to $T$. 

Since the coupling parameters are common to all temperatures,
our fixed scale approach with the $T$-integral method has several
advantages over the  
conventional approach;
(i) $T=0$ subtractions can be done by a common zero temperature simulation, 
(ii) the condition to follow the LCP is obviously satisfied, and 
(iii) the lattice scale as well as beta functions are required only at
the simulation point.  
As a result of these, the computational cost needed for $T=0$ simulations is reduced largely.
We may even borrow results of existing high precision spectrum studies at $T=0$
which are public e.g.\ on the International Lattice Data Grid \cite{ILDG}.
On the other hand, when the beta functions are not available, we need to 
perform additional $T=0$ simulations around the simulation points as in 
the case of the conventional method.

For a continuum extrapolation, we need to repeat the calculation at a couple 
of lattice spacings. 
If we adopt coupling parameters from $T=0$ spectrum studies
in which the continuum extrapolation has already been performed, we can get all
configurations for $T=0$ subtractions. 
Furthermore, because the lattice spacings in spectrum studies are usually
smaller than those used in conventional fixed-$N_t$ studies around the 
critical temperature $T_c$, the values of $N_t$ in our approach are much 
larger there than those in conventional studies.
For example, at $a \approx 0.07$ fm, $T \sim 175$ MeV is achieved by
$N_t \sim 16$. 
Therefore, for thermodynamic quantities around $T_c$, 
we can largely reduce the lattice artifacts due to large $a$ and/or small $N_t$ 
over the conventional approach,
without high computational cost for $T=0$ calculations. 
This is also a good news for phenomenological applications of the EOS, since
the temperature achieved in the relativistic heavy ion collision
at RHIC and LHC will be at most up to a few times the critical
temperature \cite{Hirano:2008hy}. 
On the other hand, calculation of the trace
anomaly around and below $T_c$ with large values of $N_t$ may require high
statistics due to large cancellations by the $T=0$ subtraction.
We note here that, as $T$ increases, $N_t$ becomes small and hence the
lattice artifact increases.  
Therefore, our approach is not suitable for
studying how the EOS approaches the Stephan-Boltzmann value in the
high $T$ limit. 

Outline of this paper is as follows.
After introducing our lattice action and the trace anomaly
in Sect.\ref{sect2}, we test our $T$-integral method in SU(3) gauge 
theory on isotropic lattice in Sect.\ref{sect3} and
on anisotropic lattice in \ref{sect3.2}. 
The last section is devoted to summary and conclusions.

\section{Lattice action}
\label{sect2}

We study the SU(3) gauge theory with the standard plaquette gauge
action on an isotropic and anisotropic lattices with the spatial
(temporal) 
lattice size $N_s$ ($N_t$) and lattice spacing $a_s$ ($a_t$). 
The lattice action is given by 
\begin{eqnarray}
S &=& 
\beta \xi_0
\sum_x\sum_{i=1}^3
\left[ 1-\frac{1}{3}\mbox{Re}\mbox{Tr} U_{i4}(x) \right]\nonumber\\
&& + \frac{\beta}{\xi_0}
\sum_x\sum_{i>j=1}^3
\left[ 1-\frac{1}{3}\mbox{Re}\mbox{Tr} U_{ij}(x) \right]\\
& \stackrel{\rm def.}{=} & 3N_s^3 N_t \, \beta \left\{ \xi_0 P_t +
\xi_0^{-1}P_s \right\}  
\end{eqnarray}
where $U_{\mu\nu}(x)$ is the plaquette in the $\mu\nu$ plane
and $\beta$ and $\xi_0$ are the bare lattice gauge coupling and bare
anisotropy parameters. 
The trace anomaly is obtained as
\begin{eqnarray}
\frac{\epsilon-3p}{T^4} &=&
\frac{N_t^3}{N_s^3 \, \xi^3} \, a_s 
\left(\frac{\partial\beta}{\partial a_s}\right)_{\!\!\xi}
\left\langle \left(\frac{\partial S}{\partial\beta}\right)_{\!\!\xi}
\right\rangle 
\label{eq:6}\\
&=&
\frac{3N_t^4}{\xi^3}
\left\langle
\left( a_s \frac{\partial \beta}{\partial a_s}
\right)_{\!\!\xi}
\left[
\left\{ \frac{1}{\xi_0} P_s + \xi_0 P_t \right\}
\right.\right.\nonumber\\
&&\left.\left. - \, \frac{\beta}{\xi_0}
\left(\frac{\partial \xi_0}{\partial \beta}\right)_{\!\!\xi}
\left\{ \frac{1}{\xi_0}P_s - \xi_0 P_t \right\}
\right]
\right\rangle \label{eq:7}
\end{eqnarray}
where $\xi=a_s/a_t$ is the renormalized anisotropy and
$a_s (\partial\beta/\partial a_s)_\xi$ is the beta function.
Note that $(\partial \xi_0/\partial \beta)_\xi = 0$ on isotropic lattices.

\section{EOS on isotropic lattice}
\label{sect3}

\begin{table}[tb]
\begin{tabular}{c|cccccccc}
\hline
set & $\beta$ & $\xi$ & $N_s$ & $N_t$ &$r_0/a_s$ & $a_s$[fm] &
$L$[fm] & $a(dg^{-2}/da)$ \\
\hline
i1 & 6.0 & 1 & 16 & 3-10 & 5.35($^{+2}_{-3}$) & 0.093 & 1.5 & -0.098172 \\
i2 & 6.0 & 1 & 24 & 3-10 &5.35($^{+2}_{-3}$) & 0.093 & 2.2 & -0.098172 \\
i3 & 6.2 & 1 & 22 & 4-13 & 7.37(3) & 0.068 & 1.5 & -0.112127 \\
\hline
a2 & 6.1 & 4 & 20 & 8-34 & 5.140(32) & 0.097 & 1.9 & -0.10704 \\
\hline
\end{tabular}
\caption{Simulation parameters on isotropic and anisotropic lattices.
On isotropic lattices, we adopt $r_0/a$ of \cite{Edwards:1997xf}, and 
the beta function of \cite{Boyd:1996bx}.
Anisotropic $r_0/a_s$ is from \cite{Matsufuru:2001cp}, while the beta
function is calculated in Sect.\ref{sect3.2}. 
The lattice scale $a_s$ and lattice size $L=N_s a_s$ are calculated
with $r_0=0.5$ fm.} 
\label{tab:para1}
\end{table}

\begin{figure}[tb]
\resizebox{75mm}{58mm}{
\includegraphics{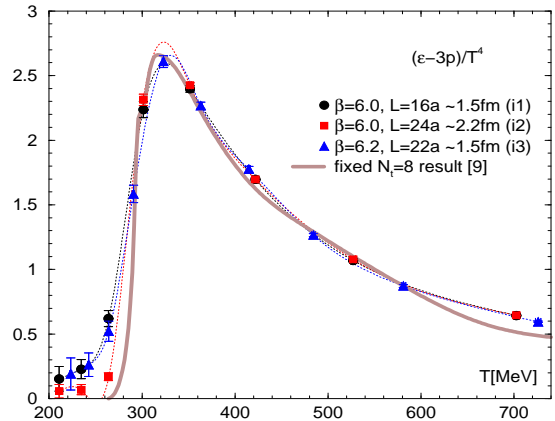}}
\caption{Trace anomaly on isotropic lattices as a function of physical
temperature.
The dotted lines are natural cubic spline interpolations.
Horizontal errors due to the lattice scale are smaller than the symbols. 
}
\label{fig:iso1}
\end{figure}

\begin{figure}[tb]
\resizebox{75mm}{58mm}{
\includegraphics{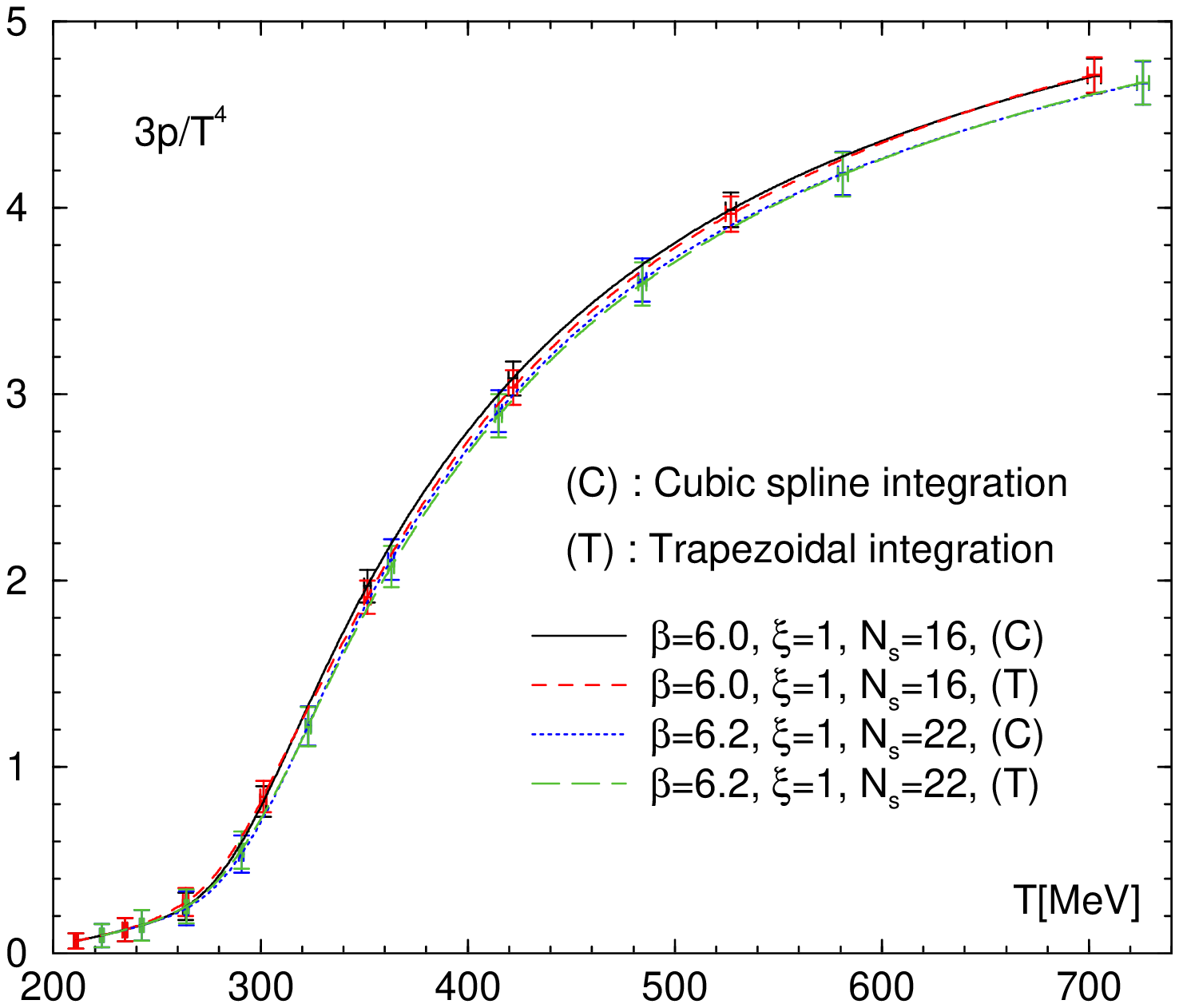}}
\resizebox{75mm}{58mm}{
\includegraphics{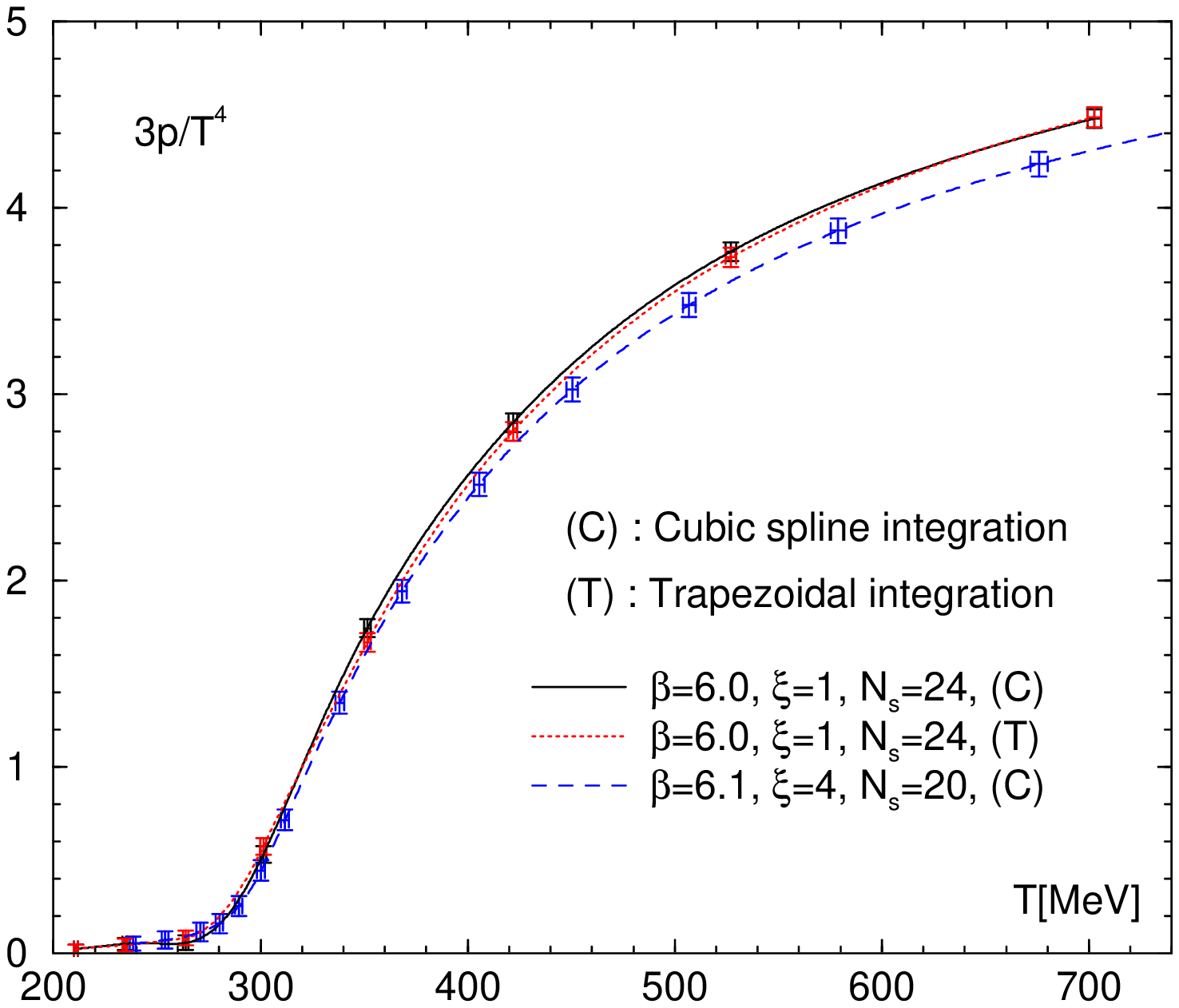}}
\caption{Comparison of the pressure using different interpolation
functions for integration. 
Statistical errors are estimated by the jackknife method
\protect\cite{Okamoto:1999hi}.} 
\label{fig:comp2}
\end{figure}

Our simulation parameters are listed in Table~\ref{tab:para1}.
On isotropic lattices, we calculate EOS on three lattices to study the
volume and lattice spacing dependence.  
The ranges of $N_t$ correspond to $T=210$--700 MeV for the sets i1 and
i2, and $T=220$--730 MeV for i3, respectively. 
The critical temperature corresponds to $N_t \sim 7$-8 for i1 and i2
and $\sim 10$ for i3, assuming $T_c\sim 290$ MeV in quenched QCD with
the Sommer scale $r_0 = 0.5$ fm. 
The set a2 will be discussed in the next section.
The zero temperature subtraction is performed with $N_t=16$ for i1 and
i2, and with $N_t=22$ for i3. 
We generate up to a few millions configurations using the
pseudo-heat-bath algorithm. 
Statistical errors are estimated by the jackknife analysis. 
Typically, bin size of a few thousands configurations are necessary
near $T_c$, while a few hundreds are sufficient off the transition
region.  

Figure~\ref{fig:iso1} shows $(\epsilon-3p)/T^4$. 
Dotted lines in the figure are the natural cubic spline interpolations. 
For comparison, we also reproduce the result of the fixed $N_t$ method at 
$N_t=8$ and  $N_s=32$ \cite{Boyd:1996bx}, for which we have rescaled 
the horizontal axis by $T_c=290$ MeV according to our choice of $r_0=0.5$ fm.
At and below $T_c$, lattice size dependence is visible among the
sets i1 ($L\approx1.5$ fm), i2 (2.2 fm) and the fixed $N_t=8$ result (2.7 fm). 
On the other hand, the lattice spacing dependence is negligible
between i1 ($a\approx0.093$ fm) and i3 (0.068 fm). 
At higher $T$, $(\epsilon-3p)/T^4$ on our three lattices show good
agreement, however they slightly deviate from the fixed $N_t$ one at
$T>600$ MeV, presumably due to the coarseness of our lattices at these temperatures.

The integration of (\ref{eq:1}) is performed numerically using the 
natural cubic spline interpolations shown in Fig.~\ref{fig:iso1}.
For the initial temperature $T_0$ of the integration, 
we linearly extrapolate the $(\epsilon-3p)/T^4$ data at a few lowest $T$'s
because the values of $(\epsilon-3p)/T^4$ at our lowest $T$ are not
exactly zero. 
In this study, we commonly take $T_0=150$ MeV as the initial
temperature which satisfies $(\epsilon - 3 p)/T^4 = 0$, and estimate
the integration from $T_0$ to the lowest $T$ by the area of the
triangle. 

We estimate the statistical error for the $T$-integration by the
jackknife analysis at each $T$ and accumulate the contributions from
different $T$'s by the error propagation \cite{Okamoto:1999hi},
because simulations at different $N_t$ are statistically independent. 
Note that the error for the lattice scale do not affect the
dimensionless quantity $p/T^4$. 
Error bars shown in the figures represent the statistical errors.

\begin{figure}[tb]
\resizebox{75mm}{58mm}{
\includegraphics{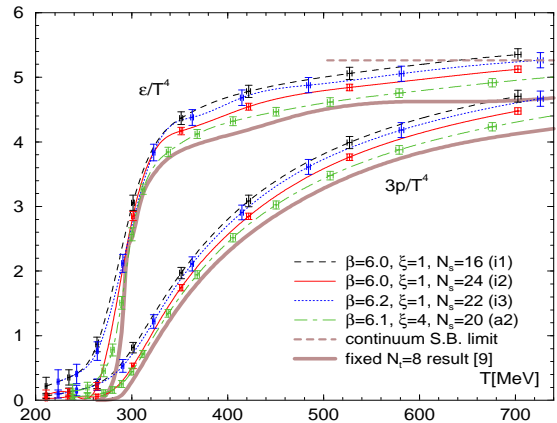}}
\caption{EOS on isotropic and anisotropic lattices.
Dashed horizontal line represents the free gas case in the continuum
(Stefan-Boltzmann limit).}  
\label{fig:iso2}
\end{figure}

To estimate the systematic error due to the interpolation ansatz, we
compare the results of $p/T^4$ 
using cubic spline interpolation and those with the trapezoidal rule
in Fig.~\ref{fig:comp2}.  
(The results on the anisotropic $\xi=4$ lattice will be discussed later.)
We find that the size of systematic errors due to the interpolation ansatz are
comparable to that of the statistical ones. 
In Fig.~\ref{fig:iso2}, we note that the natural cubic spline
interpolation curve of $(\epsilon-3p)/T^4$ for the set i2 shows small
negative values at $T\approx250$ MeV due to the nearby sharp edge at
$T\approx 260$ MeV. 
From a comparison with the results of trapezoidal interpolation for
that range, we find that the effect of this bump on the value of
$p/T^4$ at $T > 250$ MeV is 0.032. Although a negative pressure is
unphysical, because this shift in $p/T^4$ is smaller than the
statistical errors, we disregard the effects of the negative pressure
in this paper. 
%
To avoid arbitrary data handlings, we just adopt the results of
natural cubic spline interpolations as the central values in the
followings. 

In Fig.~\ref{fig:iso2}, we summarize the results of EOS. 
Results on the anisotropic lattice (a2) will be discussed later. 
The normalized energy density $\epsilon/T^4$ is calculated by
combining $p/T^4$ and $(\epsilon-3p)/T^4$.  

We find that, except for the vicinity of $T_c$, the EOS is rather
insensitive to the variation of lattice size (between $L \approx 1.5$
fm and $2.2$ fm) and the lattice spacing (between $a \approx 0.093$ fm
and 0.063 fm).  
The results of EOS agree within 10\% for our variation of lattice
parameters. 
This is in part due to the fact that $(\epsilon-3p)/T^4$ is not so
sensitive to the lattice parameters up to high 
temperatures.  
Note also that, because $(\epsilon-3p)/T^4 = 0$ in the high
temperature limit, the increasingly large lattice artifacts at large
$T$ are naturally suppressed in the $T$-integration.  
Looking at the $T$ \gsim\ $T_c$ region closer, we note a slight
tendency that both $p$ and $\epsilon$ decrease as the lattice size
(lattice spacing) becomes larger (smaller). 

Near and below $T_c$, we observe a sizable finite size effect between
$L \approx 1.5$ fm (i1) and $2.2$ fm (i2), while the effect of the
lattice spacing is quite small (i1 and i3). 
Therefore, for a reliable simulation, we need, at least, $L$ \gsim\ 2 fm. 

Our results are qualitatively consistent with the previous EOS by the
fixed $N_t$ method \cite{Boyd:1996bx}. 
Quantitatively our results are slightly above the fixed $N_t$ results.
The discrepancy can be in part understood by smaller spatial
volumes below $T_c$ and 
the small values of $N_t$ at higher $T$ in our method. 
%

\section{EOS on anisotropic lattice}
\label{sect3.2}

The anisotropic lattice with the temporal lattice finer than the
spatial one is expected to improve the resolution of $T$ 
without much increasing the computational cost.
To further test the systematic error due to the resolution of $T$, 
we perform the study with the $T$-integral method on an anisotropic lattice 
with the renormalized anisotropy $\xi = 4$. 

The simulation parameters are given as the set a2 in
Table~\ref{tab:para1}, which 
are the same as those adopted in \cite{Matsufuru:2001cp}.  
We vary $N_t=34$--8 corresponding to $T=240$--1010 MeV.
The zero temperature subtraction is performed with $N_t=80$. %
($=20\times \xi$).  
We generate up to a few millions configurations. 

We calculate the beta function $a_s (\partial\beta/\partial
a_s)_{\xi}$ by fitting the $r_0/a_s$ data
\cite{Matsufuru:2001cp,mapple} with the Allton's ansatz
\cite{Allton:1996dn}. 
\begin{equation}
a_s/r_0 = R(\beta) \cdot A\left( 1+B\hat{f}^2(\beta)+C\hat{f}^4(\beta)\right),
\end{equation}
where
\begin{eqnarray}
R(\beta) &=& \left( \frac{6b_0}{\beta}\right)^{\!-b_1/(2b_0^2)}
\exp{\left(-\,\frac{\beta}{12b_0}\right)},
\nonumber\\
\hat{f}(\beta) &=& R(\beta)/R(6.10),\;\; b_0=\frac{11}{16\pi^2},\;\;
b_1=\frac{102}{(16\pi^2)^2}. 
\nonumber
\end{eqnarray}
The best fit is achieved by $A=76.0(1.5)$, $B=0.190(32)$, and
$C=0.0204(87)$ with $\chi^2/{\rm dof}=12.6/6$.  
The beta function is now obtained by 
\begin{eqnarray}
a_s \left(\frac{\partial \beta}{\partial a_s}\right)_{\!\!\xi}
=\frac{12b_0^2\beta}{6b_1-b_0\beta}
\, \frac{1+B\hat{f}^2+C\hat{f}^4}{1+3B\hat{f}^2+5C\hat{f}^4},
\end{eqnarray}
and its value at our simulation point is given in Table~\ref{tab:para1}.
For $(\partial \xi_0/\partial \beta)_\xi$ 
we adopt the result of \cite{Klassen:1998ua}. 

\begin{figure}[tb]
\resizebox{75mm}{58mm}{
\includegraphics{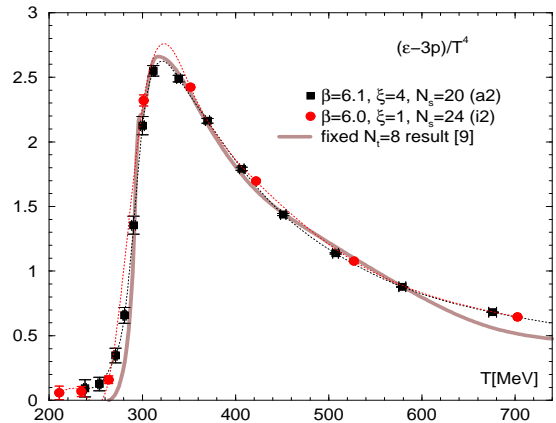}}
\caption{Trace anomaly on isotropic and anisotropic lattices with
similar spatial lattice spacing and lattice size. 
The dotted lines are natural cubic spline interpolations.
}
\label{fig:comp1}
\end{figure}

In Fig.~\ref{fig:comp1}, we compare the trace anomaly obtained on the
anisotropic lattice with that on the isotropic lattice with similar
$a_s$ and $L$ (the set i2). 
We find that the results are generally consistent with each other, while, 
due to the cruder resolution in $T$, the natural cubic spline
interpolation for the i2 lattice slightly overshoots the data on the
a2 lattice around the peak at $T\approx 320$ MeV. 
We note that the height of the peak on the a2 lattice is similar to
that of the fine i3 lattice shown in Fig.~\ref{fig:iso1} and the 
fixed $N_t=8$ result.
Similar to the case of the isotropic lattice, our result slightly
overshoots the fixed $N_t$ result below $T_c$ and at $T > 600$ MeV. 
Around the sharp lower edge at $T \approx 250$ MeV, the undershooting
of the natural cubic spline interpolation on the i2 lattice, discussed
in the previous section, is avoided by the finer data points on the a2
lattice.

The results of pressure on the a2 and i2 lattices are compared in the
lower panel of Fig.~\ref{fig:comp2}. 
While they are roughly consistent with each other, we find that the
pressure on the anisotropic lattice is slightly smaller than that on
the isotropic lattice. 
This is in part due to the smaller values of $\epsilon-3p$ around the
peak at $T\approx320$ MeV. 
We also note a slight tendency that the trace anomaly on the a2
lattice is systematically lower than that on the i2 lattice by about
$1\sigma$ in the high temperature side.  

When we consider the difference between the a2 and i2 lattices as the
systematic error due to the cruder $T$-resolution of the i2 lattice,
the systematic error is about 2-3 times larger than the estimation in
the previous section from a comparison of different interpolation
ans\"atze.  
Therefore, for an estimation of the systematic error on isotropic
lattices, it may be safer to assume a few times larger errors than
those estimated from a comparison of different interpolation
ans\"atze. 

Some part of this difference in the EOS between a2 and i2 lattices
might be explained by the difference of the lattice spacing in the
temporal direction, since lattice artifacts of thermodynamic
quantities generally have dominant contributions from the temporal
lattice spacing \cite{Namekawa:2001}.
Indeed this interpretation is supplemented by the
fact that the difference in the EOS between i1 and i3 has a similar
tendency as shown in Fig.\ref{fig:iso2}.
We reserve further investigation of this possibility for future study.

\section{Conclusions}
\label{sect4}

We proposed a fixed scale approach to investigate finite temperature
QCD on the lattice.  
To calculate EOS non-perturbatively at fixed scale, we introduced the
$T$-integral formula (\ref{eq:1}), in which $p$ is calculated as an
integration of 
the trace anomaly $\epsilon-3p$ with respect to $T$. 
The fixed scale approach with the $T$-integral method enables us to efficiently 
utilize the results of previous $T=0$ spectrum studies.
At intermediate and low temperatures, we can largely reduce the
lattice artifacts than the fixed $N_t$ approach.  
On the other hand, our approach is not suitable for studying the
approach to the high $T$ limit because of small $N_t$ there.
Also, around and below $T_c$, the large values of $N_t$  may require high
statistics due to large cancellations by the $T=0$ subtraction.

To test the $T$-integral method, we performed a series of simulations
of SU(3) gauge theory on isotropic and anisotropic lattices. 
We found that the method works quite well.
Our results of EOS for the quenched QCD on isotropic and anisotropic
lattices are summarized in Fig.~\ref{fig:iso2},  
which are qualitatively consistent with the previous results using the
conventional fixed $N_t$ approach.  

Our EOS in the high temperature region turned out to be roughly
independent of the lattice spacing, lattice volume, and the
anisotropy.   
All the results agree within about 10\% for the range of our lattice
parameters. 
With small statistical errors of less than about 2\%, we identified
small systematic shifts under the variations of lattice parameters. 
The smallness of them will be useful for precise continuum extrapolations. 
Around the critical temperature, we found that the lattice size should
be at least larger than about 2 fm. 

The fixed scale approach with the $T$-integral method is applicable to
QCD with dynamical quarks too. 
With dynamical quarks, it is more convenient to simulate isotropic
lattices, because the tuning of anisotropy parameters as well as the
determination of the factors $\partial \xi_0 / \partial\beta$ in
(\ref{eq:7}) requires quite a few works for full QCD. 
Therefore, it is important to estimate the systematic error in EOS due
to the limited resolution of $T$ on an isotropic lattice. 
From the test of quenched QCD presented in this paper, we found that
this systematic error is under control and its order of magnitude can
be correctly estimated by a comparison of different interpolation
ans\"atze, though it may be safer to introduce a factor of about 2-3
to the estimates. 

We are currently investigating EOS in $2+1$ flavor QCD with
non-perturbatively improved Wilson quarks,  
using the configurations by the CP-PACS/JLQCD Collaboration
\cite{nf2p1-wilson}, which are public on the ILDG \cite{ILDG}. 
The pseudo-critical temperature ($\sim 175$ MeV) is around $N_t \sim
16$ on the finest lattice with $a \approx 0.07$ fm. 
We are further planning to extend the study to use the $2+1$ flavor
configurations by the PACS-CS Collaboration generated just at the
physical values of the light quark masses \cite{PACS-CS}. 

TU thanks H.~Matsufuru for helpful discussions and 
$r_0/a_s$ data on the anisotropic lattice.
The simulations have been performed on supercomputers 
at RCNP, Osaka University and YITP, Kyoto University. 
This work is in part supported by Grants-in-Aid of the Japanese Ministry
of Education, Culture, Sports, Science and Technology
(Nos.~17340066, 18540253, 19549001, and 20340047). 
SE is supported by U.S.\ Department of Energy (DE-AC02-98CH10886).

\end{document}